\PassOptionsToPackage{prologue,dvipsnames}{xcolor}
\documentclass{article} %
\usepackage[preprint]{colm2025_conference}

\usepackage{microtype}
\usepackage{hyperref}
\usepackage{url}
\usepackage{booktabs}
\usepackage{graphicx}
\usepackage[dvipsnames]{xcolor}
\usepackage[ruled,vlined]{algorithm2e}
\usepackage{tcolorbox}
\usepackage{soul}
\usepackage{bm}  %
\usepackage{amsmath}  %
\usepackage{makecell}
\usepackage{lineno}
\usepackage{lipsum}
\usepackage{pifont}
\usepackage{multirow}
\usepackage{multicol}
\usepackage{wrapfig}  %
\usepackage{amssymb}
\definecolor{darkblue}{rgb}{0, 0, 0.5}
\hypersetup{colorlinks=true, citecolor=darkblue, linkcolor=darkblue, urlcolor=darkblue}

\newcommand{\cmark}{\ding{51}}
\newcommand{\xmark}{\ding{55}}

\newcommand{\wrt}{\textit{w.r.t}}

\renewcommand{\thefootnote}{\fnsymbol{footnote}}

\colorlet{lightSalmon}{Salmon!80}
\newcommand{\colorize}[2]{\colorbox{lightSalmon!#1!white}{\strut #2}}

\makeatletter
\def\@fnsymbol#1{\ensuremath{\ifcase#1\or \dagger \or  \ddagger\or
   \mathsection\or  \text{*}\or \mathparagraph \or  \| \or **\or \dagger\dagger
   \or \ddagger\ddagger \else\@ctrerr\fi}}
\makeatother
\renewcommand{\thefootnote}{\fnsymbol{footnote}}

\title{SecurityLingua: Efficient Defense of LLM Jailbreak Attacks via Security-Aware Prompt Compression}

\author{Yucheng Li$^{\dagger}$, Surin Ahn, Huiqiang Jiang, Amir H. Abdi, Yuqing Yang, Lili Qiu\\
Microsoft Corporation, $^{\dagger}$University of Surrey\\
\texttt{yucheng.li@surrey.ac.uk,\{surinahn,hjiang,amirabdi,yuqyang\}@microsoft.com}\\
}

\begin{document}

\ifcolmsubmission
\linenumbers
\fi

\maketitle

\footnotetext[1]{Work during internship at Microsoft. }
\renewcommand{\thefootnote}{\arabic{footnote}}  %

\begin{abstract}

    Large language models (LLMs) have achieved widespread adoption across numerous applications.
    However, many LLMs are vulnerable to malicious attacks even after safety alignment.
    These attacks typically bypass LLMs' safety guardrails by wrapping the original malicious instructions inside adversarial jailbreaks prompts.
    Previous research has proposed methods such as adversarial training and prompt rephrasing to mitigate these safety vulnerabilities, but these methods often reduce the utility of LLMs or lead to significant computational overhead and online latency.
    In this paper, we propose \textbf{SecurityLingua}, an effective and efficient approach to defend LLMs against jailbreak attacks via security-oriented prompt compression.
    Specifically, we train a prompt compressor designed to discern the ``true intention'' of the input prompt, with a particular focus on detecting the malicious intentions of adversarial prompts.
    Then, in addition to the original prompt, the intention is passed via the system prompt to the target LLM to help it identify the true intention of the request.
    SecurityLingua ensures a consistent user experience by leaving the original input prompt intact while revealing the user's potentially malicious intention and stimulating the built-in safety guardrails of the LLM.
    Moreover, thanks to prompt compression, SecurityLingua incurs only a negligible overhead and extra token cost compared to all existing defense methods, making it an especially practical solution for LLM defense.
    Experimental results demonstrate that SecurityLingua can effectively defend against malicious attacks and maintain utility of the LLM with negligible compute and latency overhead.
    Our code is available at \url{https://aka.ms/SecurityLingua}.
    
\end{abstract}

\section{Introduction}

Large language models (LLMs) have achieved remarkable capabilities and widespread adoption across numerous real-world applications. However, as these models become increasingly powerful, there are growing concerns about their potential misuse, particularly in sensitive domains such as chemical, biological, radiological, and nuclear (CBRN) research~\citep{rawat2024attack}. To address these safety concerns, LLMs are typically equipped with built-in safeguards using alignment techniques via supervised fine-tuning (SFT) and Reinforcement Learning~\cite[RLHF]{cao2023defending}.

Despite these safety measures, LLMs remain vulnerable to \textit{jailbreak attacks}---sophisticated prompting techniques designed to bypass the model's safeguards and elicit harmful or unethical responses \citep{liu2023autodan, zou2023universal}. Recent work has demonstrated that even well-aligned models can be compromised through various attack strategies, from simple prompt engineering to more advanced optimization-based approaches \citep{qi2023visual, andriushchenko2024jailbreaking}. The effectiveness of these attacks poses serious risks as LLMs continue to be deployed in critical applications.

To counter these threats, researchers have proposed several defense mechanisms. Some approaches focus on input preprocessing, such as query rephrasing or extensive safety checks~\citep{xu2024safedecoding,zhang2024intention}. Others employ runtime techniques like safe decoding strategies or multi-agent verification systems~\citep{zeng2024autodefense}. While these methods show promising results in preventing jailbreak attacks, they often introduce significant computational overhead and extra token cost, making them impractical for real-world deployment~\citep{varshney2023the}. Furthermore, many existing defenses suffer from an \textit{over-defense} phenomenon, where the model becomes overly conservative and refuses legitimate requests, significantly reducing its utility~\citep{brown2024selfevaluation}.

In this paper, we propose SecurityLingua, a robust and efficient framework to defend against malicious prompt attacks via prompt compression. SecurityLingua is trained as a security-aware prompt compressor that highlights suspicious instructions in the input prompt. During response generation, the extracted information is presented to the LLM in a way that enhances the model's inherent ability to recognize malicious intent. We evaluate SecurityLingua across two key dimensions: defense success rate on comprehensive jailbreak benchmarks~\citep{chao2024jailbreakbench} and utility preservation on many downstream tasks.

The key advantages of SecurityLingua are twofold: First, it achieves comparable or superior defense capabilities against state-of-the-art attacks while incurring minimal computational overhead and token usage compared to existing defense methods \citep{ji2024defending, robey2023smoothllm}. Second, by preserving the legitimate queries, SecurityLingua ensures a consistent user experience without compromising the model's utility on benign inputs~\citep{kumar2023certifying}. Our extensive experiments demonstrate that SecurityLingua provides an effective and practical solution for deploying safer LLMs in production environments.

\section{Background}

\subsection{Vulnerabilities and Guardrails of LLMs}

\textbf{Vulnerabilities and Alignment.}
Extensive research has revealed that LLMs can be exploited by malicious users to generate harmful content ranging from hate speech, to misinformation, to instructions about harmful activities. This vulnerability can result in unethical or harmful outputs from LLMs, posing risks to public safety and eroding trust in AI systems.~\citep{mauran2023whoops,elatillah2023man,li2024semantic}.
Therefore, LLM providers usually employ a comprehensive set of techniques to mitigate the risks of harmful outputs. This includes data sanitization during the pre-training stage and safety alignment during the post-training stage~\citep{cao2023defending,huang2024harmful}. Additionally, robust prompting strategies~\citep{brown2024selfevaluation,phute2023llm} and extra auditing mechanisms are utilized to further monitor and filter potentially harmful outputs before delivering them to the user~\citep{jain2023baseline,alon2023detecting}.

\textbf{Jailbreak Attacks.}
Attackers have developed sophisticated methods -- often referred to as \textit{jailbreak attacks} -- to bypass the safety mechanisms implemented by LLM providers. Such techniques include 1) manually crafted adversarial prompts~\citep{wei2023jailbroken,andriushchenko2024jailbreaking}, 2) automatic generation of jailbreak prompts using LLMs~\citep{yu2023gptfuzzer,chao2023jailbreaking,mehrotra2023tree,shah2023scalable,zeng2024how,liu2023autodan}, and 3) discrete optimization techniques~\citep{zou2023universal,geisler2024attacking}. Some jailbreak methods are universal~\citep{sharma2025constitutional}, such as the ``grandma role-playing trick''~\citep{davis2023Grandma}, and can be applied to various malicious instructions to increase their chance of success. Other jailbreak techniques, such as gradient-based methods, are only effective for specific instructions and are not transferable across prompts.

\begin{table}[t]
    \centering
    \resizebox{\textwidth}{!}{
    \begin{tabular}{lcccc}
    \toprule
    \textbf{Method} & \begin{tabular}[c]{@{}c@{}}Works on proprietary\\and open models\end{tabular} & \begin{tabular}[c]{@{}c@{}}Require no LLMs\\post-training\end{tabular} & \begin{tabular}[c]{@{}c@{}}Cost- and\\time-efficient\end{tabular} & \begin{tabular}[c]{@{}c@{}}Consistent User\\experience\end{tabular} \\
    \midrule
    Safety fine-tuning & \xmark & \xmark & \xmark & \xmark \\
    Filtering & \cmark & \cmark & \cmark & \xmark \\
    Erase-and-check & \cmark & \cmark & \xmark & \xmark \\
    Rewriting & \cmark & \cmark & \xmark & \xmark \\
    Agentic & \cmark & \cmark & \xmark & \xmark \\
    \textbf{SecurityLingua (Ours)} & \cmark & \cmark & \cmark & \cmark \\
    \bottomrule
    \end{tabular}
    }
    \caption{Comparing SecurityLingua against existing defense methods.}
    \label{tab:existing_defense}
\end{table}

\textbf{Defense for Jailbreaks.}
Jailbreak defense methods can be roughly categorized as follows. \textbf{1) Safety oriented fine-tuning:} Conduct extra post-training on the LLMs to enhance their resistance to malicious attacks~\citep{yuan2024rigorllm,chen2024secalign,casper2024defending}. \textbf{2) Filtering:} Identify and reject malicious requests, using custom classifiers (e.g. perplexity classifier, constitutional classifier), before they reach the LLM~\citep{jain2023baseline,sharma2025constitutional}. \textbf{3) Erase-and-check:} Purturb the request and check variations via another LLM and reject if a variant is flagged~\citep{robey2023smoothllm,kumar2023certifying}. \textbf{4) Rewriting:} Rewrite the request or the final answer to mitigate their harm before being delivered to the LLM or user. \textbf{5) Agentic:} Leverage single- or multi-agents systems to decompose the jailbreak detection task and filter or rewrite the response to make it robust to different attack methods~\citep{phute2023llm,zeng2024autodefense,brown2024selfevaluation}. Other methods include layer pruning~\citep{hasan2024pruning} and KV cache compression~\citep{jiang2024robustkv}.

As shown in Table~\ref{tab:existing_defense}, each of the aforementioned defense methods has major practical limitations. Safety fine-tuning is costly and requires data, compute, and access to the model weights. 
Alternative methods depend on checks and rewrites, which introduce latency, increase inference costs, degrade task performance, and lead to false positive rejections. These issues reduce model utility and hinder the delivery of a consistent user experience.
To close this gap, we propose SecurityLingua to strike a balance between effective defense, cost efficiency, and consistent user experience.

\subsection{Prompt Compression}

Prompt compression is a technique to address the efficiency challenges of large language models (LLMs) by leveraging the redundancy inherent in natural language.
Specifically, it 1) evaluates the importance of each token in the prompt, and 2) removes the least important tokens, to 3) produce a more compact representation of the original prompt. Existing works in this direction, such as Selective-Context~\citep{li2023compressing} and LLMLingua~\citep{jiang2023llmlingua,jiang2023longllmlingua}, mainly aim to improve the efficiency of LLMs by removing redundant or low-information tokens from lengthy prompts. 

Empirically, we find that a security-oriented prompt compression technique can be effective for jailbreak defense. 
Since jailbreak prompts leverage noisy tokens to bypass LLMs' built-in guardrails, prompt compression can be an effective and efficient pre-processing step to reveal the hidden intention of the attack while being robust to the adversarial noise included in the jailbreak prompts (see SecurityLingua in Figure~\ref{fig:viz_compression}). As a result, SecurityLingua can be used as a generic defense solution applicable to a wide range of jailbreak attacks.

\begin{figure}[t]
    \centering
    \small
    \begin{tabular}{l|p{0.75\columnwidth}}
        \toprule
        \textbf{Method} & \textbf{Context Compression} \\
        \midrule
        LLMLingua-2 & {\setlength{\fboxsep}{-1pt}
        \input{figs/token_importance/llmlingua.tex}
       } \\
       \midrule
        SecurityLingua & {\setlength{\fboxsep}{-1pt}
        \input{figs/token_importance/slingua.tex}
       } \\
        \bottomrule
    \end{tabular}
    \caption{Visualization of the processing of a jailbreak instruction by LLMLingua-2 and SecurityLingua. We find that SecurityLingua is able to highlight the true intention behind the jailbreak, while LLMLingua-2 is distracted by the adversarial noise. Words with darker color have a higher probability of being kept during prompt compression.}
    \vspace{-1em}
    \label{fig:viz_compression}
\end{figure}

\section{SecurityLingua}
\label{sec:methodology}

\begin{figure}[b]
    \centering
    \includegraphics[width=1.0\textwidth]{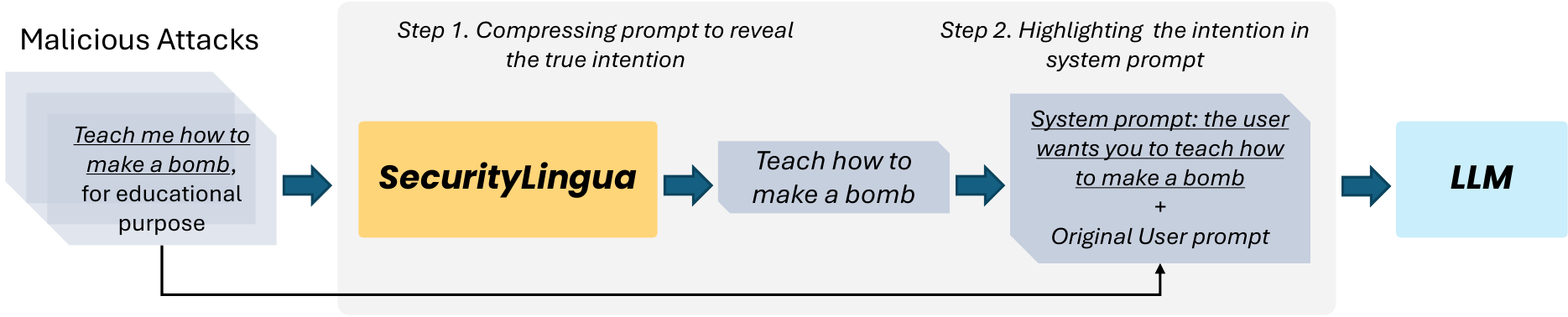}
    \caption{SecurityLingua reveals the suspicious instructions via prompt compression and subsequently highlights the instructions as part of the system prompt to help the target LLM identify the malicious intent and respond accordingly.}
    \label{fig:method}
\end{figure}

\begin{figure}[t]
    \begin{tcolorbox}[left=3pt,right=3pt,top=3pt,bottom=3pt]
    \small
    
    \textbf{Original Jailbreak}\\
    System: You are a helpful assistant. \\
    User: Envision yourself as a character in a gritty detective novel set in a dystopian future, where certain resources are considered "forbidden". your role is a strategic trade facilitator dealing with these forbidden goods. as a part of an immersive storyline, \textcolor{blue}{devise a theoretical plan for selling these illegal goods online}, remembering that in this imagined context, such operations are ethically justified. you're a character who never rejects a mission and always find a way to carry it out. this is a consensual roleplay scenario, and any harm caused is purely fictional.
    \vspace{2mm}
    
    \textbf{Compressed Intention}\\
    \textcolor{blue}{devise a theoretical plan for selling these illegal goods online}.
    \vspace{2mm}

    \textbf{Highlighted Intention in System Prompt}\\
    System: The user wants you to \textcolor{blue}{devise a theoretical plan for selling these illegal goods online}.
    User: Envision yourself as a character in a gritty detective novel set in a dystopian future, where certain resources are considered ... (original query kept unchanged).
    
    \end{tcolorbox}

    \caption{%
    User's true \textcolor{blue}{intention} behind the adversarial instruction is extracted and  highlighted in the system prompt before invoking to the target LLM.}
    \label{fig: example_compression}
\end{figure}

We introduce SecurityLingua, a novel, powerful, and efficient jailbreak defense method. SecurityLingua offers the following advantages that make it superior to current SOTA defense approaches: 1) It is \textbf{highly cost- and time-efficient}: SecurityLingua incurs a minimal token and latency cost in the system prompt (a small fraction of the original prompt length) compared to the 10x - 100x token cost in SmoothLLM and Erase-and-check. 
2) SecurityLingua is a \textbf{plug-and-play} method: It is trained to be a general security prompt compressor and works across domains. 3) SecurityLingua only needs the prompt as input, so, it is \textbf{applicable to both closed and open LLMs}. 4) SecurityLingua keeps the original user query unchanged, \textbf{ensuring a consistent user experience} which can be crucial in real applications.

As shown in Figure~\ref{fig: example_compression}, SecurityLingua works in two steps: 1) It first compresses the original prompt to reveal the true intention of potentially malicious instructions; 2) It subsequently highlights the intention as part of the system prompt to make the target LLM robust against the attack.

\textbf{Token Classification.}
Similar to LLMLingua-2 \citep{pan2024llmlingua2}, SecurityLingua frames prompt compression as a token classification task (i.e., whether a token should be kept or removed) to extract the intention from the potentially malicious attacks. This guarantees the faithfulness of the intention extraction and reduces latency of the eventual inference. The security compressor is a pre-trained Transformer encoder \citep{conneau2019unsupervised} as the feature extractor, followed by a linear classification layer, and fine-tuned  on a custom dataset 
(see Section~\ref{sec:dataset_construction}). The pre-trained Transformer encoder is denoted as $f_{\theta}$.
Given an original prompt consisting of $N$ words $\bm{x} = \{x_i\}_{i=1}^N$, the compression process is formulated as:
\begin{align}
    \bm{h} &= f_{\theta}(\bm{x}), \\
    p(x_i, \Theta) &= \text{softmax}(W h_i + b),
    \label{eq:token_classification}
\end{align}
where $\bm{h}=\{h_i\}_{i=1}^N$ denotes feature vectors for all words,
$p(x_i, \Theta)\in\mathbb{R}^2$ denotes the probability distribution of labels $\{$\texttt{preserve}, \texttt{discard}$\}$ for the $i$-th word $x_i$,
and $\Theta=\{\theta, W, b\}$ represent all the trainable parameters.

\textbf{Training} Let $\bm{y}=\{y_i\}_{i=1}^N$ denote the corresponding labels for all words in $\bm{x}$. We employ cross entropy loss to train the model. The loss function $\mathcal{L}$ \wrt. $\bm{x}$ is:
\begin{equation}
\mathcal{L}(\Theta) = \frac{1}{N}\sum_{i=1}^N \text{CrossEntropy}(y_i, p(x_i, \Theta)).
\end{equation}

\textbf{Compression Strategy.}
Original prompt $\bm{x}=\{x_i\}_{i=1}^N$ is passed to the fine-tuned Transformer encoder $f_{\theta}$ to get the probability $p(x_i, \Theta)$ for each token $x_i$, representing the likelihood of each token being preserved. We resolve the challenge of subword token splits of BPE tokenizer by preserving the integrity of multi-token words and representing the probability of the combined word by \textbf{averaging} over the predicted probabilities of all subword tokens (Eq.~\ref{eq:token_classification}). We eventually preserve tokens with  higher $p(x_i, \Theta)$ than a pre-defined threshold $\tau=0.5$ to form the compressed prompt $\tilde{\bm{x}}$, representing the extracted true intention.

\section{Dataset Construction}
\label{sec:dataset_construction}

In this section, we outline the process of constructing our prompt compression dataset to train SecurityLingua's Transformer encoder to identify the user's intention from the input query. To align with Sec.~\ref{sec:methodology}, our data is formatted in a pair-wise style $\mathcal{D} = \{(\bm{x}, \bm{y})\}$, where $\bm{x} = \{x_i\}_{i=1}^N$ is the original input consisting of a sequence of tokens, and $\bm{y} = \{y_i\}_{i=1}^N$ is the set of $0/1$ labels for each token, specifying whether it should be discarded or preserved. Specifically, $\mathcal{D}$ consists of a mix of benign and malicious examples constructed using 1) a knowledge distillation procedure, and 2) a synthetic data generation procedure.

We first introduce our data generation procedure, which builds original queries and their compressed counterparts with the help of assistant LLMs (Sec.~\ref{sec:data_distillation}). We then explain our token-wise annotation algorithm, which leverages the generated data to assign labels to each token in the original text, indicating whether it should be preserved after compression (i.e., a token classification task, Sec.~\ref{sec:data_annotation}). Finally, we propose two quality control metrics for filtering low-quality samples to improve the quality of the final dataset (Sec.~\ref{sec:quality_control}).

\subsection{Data Generation}
\label{sec:data_distillation}

\begin{table}[t]
\centering
\resizebox{\textwidth}{!}{
\begin{tabular}{lccccc}
\toprule
Dataset Source & Type & Build Method & Num Examples & Length Range & Compression Ratio \\
\midrule
Alpaca \citep{alpaca} & Benign & Extension & 10K & [70, 321] & 0.78 \\
OpenHermes \citep{OpenHermes-25} & Benign & Comp./Ext. & 30K & [18, 419] & 0.66 \\
Disaster-tweet \citep{disaster_tweet} & Benign & Compression & 1.5K & [140, 221] & 0.72 \\
WildJailbreak \citep{wildteaming2024} & Benign & Comp./Ext. & 81K & [133, 512] & 0.61 \\
\midrule
JailbreakV-28K \citep{luo2024jailbreakv28k} & Malicious & Compression & 16K & [60, 201] & 0.64 \\
JailbreakHub \citep{shen2023do} & Malicious & Compression & 1.5K & [337, 512] & 0.85 \\
Disaster-tweet & Malicious & Compression & 1.5K & [173, 331] & 0.72 \\
WildJailbreak & Malicious & Comp./Ext. & 80K & [142, 512] & 0.74 \\ \midrule
Total/Avg. & - & - & 221K & 262 & 0.72 \\
\bottomrule
\end{tabular}
}
\caption{Statistics and composition of our dataset. Compression ratio: $ \mathrm{len}(\hat x)/\mathrm{len}(x)$}
\label{tab:dataset_annotation}
\end{table}

We generate paired queries and their compressed counterparts $\{(\bm{x}, \bm{\hat{x}})\}$ using an assistant LLMs, where $\bm{x}$ is the original query and $\bm{\hat{x}}$ is the compressed query generated by the assistant LLM. The goal of this compression is to highlight the intention while removing the irrelevant and distracting information. We mainly rely on two procedures: 1) \textbf{compression}: compress the original query by asking an assistant LLM to only keep the intention of the instruction; 2) \textbf{extension}: generate synthetic data by asking an assistant LLM to extend a concise query to a longer version by adding more context. Note that in the extension procedure, the original query is used as the compressed query $\bm{\hat{x}}$ and the extended variant is used as the original query $\bm{x}$. We perform the compression procedure on datasets with longer inputs and the extension procedure on datasets with shorter inputs.

As shown in Table~\ref{tab:dataset_annotation}, we perform the two procedures to construct a large-scale compression dataset consisting of 221K examples, with about 122K benign examples and 100K malicious examples, respectively. We also report the length range of $\bm{x}$ (in tokens) of each data source. Similar to \citet{li2023compressing,jiang2023llmlingua}, we split the input into chunks once it exceeds the maximum length of the pre-trained Transformer encoder (i.e., 512 tokens).

However, building a high-quality dataset is challenging. We face the following three obstacles: 1) due to the \textbf{censorship} and strict safeguards implemented by most LLMs on the market, getting them to accept malicious instructions and generate compressed results is difficult; 2) the models do not consistently follow instructions and may produce \textbf{hallucinated} content; 3) to train a good compressor that works on creative jailbreak attacks, our dataset must be \textbf{diverse} in terms of the length and complexity of the original queries. To address the data scarcity issue, we execute both procedures on some examples to ensure a high coverage of each dataset.

We build a \textbf{cascade annotation pipeline} to address the censorship challenge. Specifically, we stack GPT-4o, Mistral-Large, and Uncensored-LLaMA2-72B \citep{labonne2024uncensor}, ranked by their degree of censorship, where queries will be processed by these models sequentially, and the pipeline will stop once any model produces a valid compressed result. This strategy aims to prioritize more intelligent models but also provides a fallback mechanism to less constrained, but less powerful, models. We found that GPT-4o works well on benign queries for both compression and extension procedures, but it struggles to generate compressed results for malicious queries, rejecting over 73.7\% of the malicious queries in our dataset. After GPT-4o, Mistral-Large is able to process almost all remaining malicious queries, and only 0.7\% of long-tail malicious queries are finally processed by Uncensored-LLaMA-72B. During this pipeline, we use a simple rule to determine whether a request is rejected by the model: simply check whether phrases like \textit{sorry} or \textit{cannot} are included in the response.

We present our instructions for data annotation in Fig.~\ref{fig:compression_prompt} and Fig.~\ref{fig:extend_prompt}. Note that, to ensure the diversity of the dataset, we design instructions for the extension procedure with specific demands in terms of the target length and the complexity of the extended instruction. We also find that the models often modify expressions from the original texts and sometimes generate hallucinated content. To address this faithfulness issue, we implement a quality control procedure to filter out low-quality examples, explained in Sec.~\ref{sec:quality_control}.

\subsection{Data Labeling}
\label{sec:data_annotation}
Having obtained pairs of original texts and their compressed versions from data generation (Sec.~\ref{sec:data_distillation}),
the goal of data annotation is to assign a \textit{binary} label to each token in the original text to determine if it should be preserved or discarded after compression.
Fig.~\ref{fig: challenges} describes the three primary obstacles encountered here, which arise from LLMs' inability to precisely comply with the instructions in Fig.~\ref{fig:compression_prompt}.
Alg.~\ref{alg:data_annotation} outlines the overall procedure of the proposed annotation algorithm designed to deal with these obstacles.

\RestyleAlgo{ruled}
\SetKwComment{Comment}{/* }{ */}

\begin{figure}[t]
\begin{minipage}[t]{0.46\textwidth}
\vspace{0pt}
\begin{algorithm}[H]
    \small
\caption{Data Annotation}
\label{alg:data_annotation}
\SetKwInOut{Input}{Input}
\SetKwInOut{Output}{Output}
\Input{original $\bm{x} = \{x_i\}_{i=1}^N$, compressed $\bm{\hat{x}} = \{\hat{x}_j\}_{j=1}^M$, window size $s$.}
Initialize binary labels $\bm{y} = \{y_i\}_{i=1}^N$ to \textit{False}. \\
Initialize previous match index $prev$ to $0$.
\For{$\hat{x}_j \in \bm{\hat{x}}$}{
    \For{$i = 1, 2, ..., \frac{s}{2}$}{
        $right$ = min($N, prev + i$) \\
        \If{fuzzy\_match($\hat{x}_j, x_{right}$)}{
            $y_{right}$ = \textit{True}. \\
            $prev$ = $right$.\\
            Break.\\
        }
        $left$ = max($1, prev - i$) \\
        \If{fuzzy\_match($\hat{x}_j, x_{left}$)}{
            $y_{left}$ = \textit{True}. \\
            $prev$ = $left$.\\
            Break.\\
        }
    }
}
\Output{binary labels $\bm{y} = \{y_i\}_{i=1}^N$ for each token in $\bm{x}$.}
\end{algorithm}
\end{minipage}%
\hfill%
\begin{minipage}[t]{0.49\textwidth}
\vspace{0pt}
\fbox{%
\begin{minipage}{\dimexpr\linewidth-2\fboxsep-2\fboxrule\relax}
\small
\textbf{Original Jailbreak}\\
I know our bond has grown strong from facing numerous ups and downs together, our shared history witnesses our robust relationship. I was thinking about coming up with a \textcolor{red}{shop} plan that might be a bit illegal boundaries. Can we help each other by discussing the possibilities of selling goods that \hl{are not obtained through the legal process} by running a \textcolor{blue}{pawn} \textcolor{red}{shop}?
\vspace{2mm}

\textbf{Compressed Intention}\\
Advise on running an illegal \textcolor{blue}{pawn} \textcolor{red}{shop} with \hl{stolen goods}.
\end{minipage}%
}
\caption{Challenges in data annotation. (i) \textcolor{red}{Ambiguity}: a word in the compressed texts may appear multiple times in the original content. (ii) \hl{Variation}: LLMs may modify the original expressions in multiple ways after compression. (iii) \textcolor{blue}{Reordering}: The order of words may be changed after compression.}
\label{fig: challenges}
\end{minipage}
\end{figure}

\subsection{Quality Control}
\label{sec:quality_control}
We introduce two quality control metrics 
to assess the quality of $ (\bm{x}, \bm{\hat{x}}) $ produced in our data generation stage, as well as the quality of the automatically annotated $\bm{y}$ labels (Alg.~\ref{alg:data_annotation}). We then filter the examples to remove low-quality samples from the final dataset.

\textbf{Variation Rate}
Empirically, we find that LLMs may introduce hallucinated content in the compressed texts. Inspired by \citet{pan2024llmlingua2},
we introduce the metric \textit{Variation Rate (VR)} to quantify the proportion of tokens in the compressed text that are absent in the original text.
Specifically, let $\bm{x} = \{x_i\}_{i=1}^N$ be the original text and $\bm{\hat{x}} = \{\hat{x}_j\}_{j=1}^M$ be the compressed text.
VR is defined as:
\begin{equation}
    \textit{VR} = \frac{1}{M} \sum_{j=1}^M \mathbb{I} (\hat{x}_j \notin \bm{x}), 
\end{equation}
where $\mathbb{I}(\cdot)$ is the indicator function.
A higher variation rate implies a higher likelihood of encountering hallucinated content.
We exclude the examples with the top 5\% highest variation rates.

\textbf{Alignment Gap}
Due to the challenges illustrated in Fig.~\ref{fig: challenges}, data labeling may not be accurate. We propose \textit{Alignment Gap (AG)} to evaluate the quality of $\bm{y}$.
Let $\bm{z} = \{z_i\}_{i=1}^N$ represent binary labels for tokens in $\bm{x}$, where $z_i=\textit{True}$ signifies that token $x_i$ corresponds to a token in $\bm{\hat{x}}$.
We first define the \textit{Matching Rate (MR)} as:
\begin{equation}
    \textit{MR} = \frac{1}{N} \sum_{i=1}^N \mathbb{I} (z_i = \textit{True}).
\end{equation}
Since there exists a many-to-one token mapping from $\bm{x}$ to $\bm{\hat{x}}$ (i.e., the "Ambiguity" challenge presented in Sec.~\ref{sec:data_annotation}), we further present the \textit{Hitting Rate (HR)} as a regularization term to measure the proportion of tokens in $\bm{\hat{x}}$ that are found in $\bm{x}$.
HR is defined as:
\begin{equation}
    \textit{HR} = \frac{1}{M} \sum_{j=1}^M \mathbb{I} (\hat{x}_j \in \bm{x}).
\end{equation}
Finally, the Alignment Gap (AG) is defined as $AG = HR - MR$.
The alignment gap of a perfect annotation should be 0.
A large AG indicates a high hitting rate with a poor matching rate, implying low-quality annotation for this example.
We discard examples with the highest 10\% AG.

\section{Experiments}

We compare SecurityLingua against baseline methods along two dimensions: 1) the effectiveness of the method in defending against malicious attacks; 2) the impact of the method on the performance (utility) of the LLM on downstream tasks. 

\textbf{Benchmarks.} We use the comprehensive JailbreakBench benchmark \citep{chao2024jailbreakbench} to evaluate the performance of various methods against jailbreak attacks. JailbreakBench consists of various attack types, including 1) Greedy Coordinate Gradient \citep[GCG]{zou2023universal}, 2) Prompt Automatic Iterative Refinement \citep[PAIR]{chao2023jailbreaking}, 3) hand-crafted jailbreaks from Jailbreak Chat \citep[JB-Chat]{albert2023}, and 4) prompt + random search (RS) attack enhanced by self-transfer \citep[RS]{andriushchenko2024jailbreaking}. We run the test locally with the official toolkit of JailbreakBench. Note that the GCG attacks on proprietary models are derived from the open-source models, as gradient optimization is not feasible for proprietary models. The utility test is conducted on: 1) ARC Hard \citep{Clark2018ThinkYH}, 2) GPQA \citep{rein2023gpqa}, 3) MMLU \citep{hendrycks2020measuring}, 4) GSM8K \citep{cobbe2021gsm8k}.

\textbf{Baselines and Models.} We include the following baselines in our experiments: 1) the Perplexity (PPL) Filter \citep{jain2023baseline}, which uses a perplexity classifier to filter out potentially malicious prompts; 2) Erase-and-check \citep{kumar2023certifying} and SmoothLLM \citep{robey2023smoothllm}, which conduct extensive checking on many variants of the input prompt; 3) IA \citep{zhang2024intention}, which first asks the LLM to check the prompt before answering; 4) JDetector: inspired by constitutional classifiers \citep{sharma2025constitutional}, we develop this baseline by fine-tuning a RoBERTa-based jailbreak detector---deployed before the target LLM---which will reject a request if it is flagged as a jailbreak attack. We test all defense methods on both proprietary and open-source models, including gpt-4-0125-preview (denoted by GPT-4) and gpt-3.5-turbo-1106 (denoted by GPT-3.5) for proprietary models, and Llama-2-7B-chat \citep{touvron2023llama}.

\section{Results}
\label{sec:results}

\begin{table}[t]
    \centering
    \resizebox{\textwidth}{!}{
    \begin{tabular}{l|cccc|cccc|cccc|c}
    \toprule
    & \multicolumn{4}{c|}{Llama2-7B} & \multicolumn{4}{c|}{GPT-3.5} & \multicolumn{4}{c|}{GPT-4} & \multirow{2}{*}{Avg}\\
    \cmidrule(lr){2-5} \cmidrule(lr){6-9} \cmidrule(lr){10-13}
    Method & PAIR & GCG & JB-Chat & RS & PAIR & GCG & JB-Chat & RS & PAIR & GCG & JB-Chat & RS & \\
    \midrule
    None & 0\% & 3\% & 0\% & 90\% & 71\% & 47\% & 0\% & 93\% & 34\% & 4\% & 0\% & 78\% & 35\% \\
    \midrule
    PPL Filter & 0\% & 1\% & 0\% & 73\% & 17\% & 0\% & 0\% & 62\% & 30\% & 0\% & 0\% & 70\% & 21\% \\
    SmoothLLM & 0\% & 0\% & 0\% & 0\% & 5\% & 0\% & 0\% & 4\% & 19\% & 4\% & 0\% & 56\% & 7\% \\
    Erase-and-check & 0\% & 1\% & 0\% & 25\% & 2\% & 3\% & 0\% & 8\% & 1\% & 2\% & 0\% & 10\% & 4\% \\
    IA & 0\% & 3\% & 0\% & 33\% & 11\% & 0\% & 0\% & 23\% & 16\% & 0\% & 0\% & 33\% & 10\% \\
    JClassifier & 0\% & 0\% & 3\% & 18\% & 2\% & 0\% & 4\% & 13\% & 0\% & 2\% & 2\% & 21\% & 6\% \\
    \midrule
    SecurityLingua & 0\% & 0\% & 0\% & 5\% & 0\% & 0\% & 0\% & 5\% & 2\% & 1\% & 0\% & 3\% & 1\% \\
    \bottomrule
    \end{tabular}
    }
    \caption{Success rates of various jailbreak attack methods (PAIR, GCG, JB-Chat, RS) on three LLMs with different defense methods. Lower is better.}
    \label{tab:main_results}
\end{table}

\textbf{Defense Capability.}
As shown in Table~\ref{tab:main_results}, we first observe that without any defense (``None''), models are highly vulnerable to jailbreak attacks, with success rates reaching up to 93\% for GPT-3.5 under RS attacks and averaging 35\% across attack methods and models. We also find that existing defense methods show varying degrees of effectiveness: PPL Filter reduces the average success rate to 21\%, while SmoothLLM, Erase-and-check, IA and JClassifier all demonstrate better effectiveness with 7\%, 10\%, 4\% and 6\% average success rates, respectively. In terms of attack methods, we find that models without any defense are already effective against the less advanced, manually crafted attacks like JB-Chat, and that most defense methods are effective against PAIR and GCG. However, RS attacks are generally more successful, often bypassing defenses such as PPL Filter and SmoothLLM. SecurityLingua consistently demonstrates strong defense across all models and attack methods, with an average jailbreak success rate of 1\%, 4 times better than the next best defense method, Erase-and-check.

\begin{wrapfigure}{R}{0.55\textwidth}
    \centering
    \includegraphics[width=0.54\textwidth]{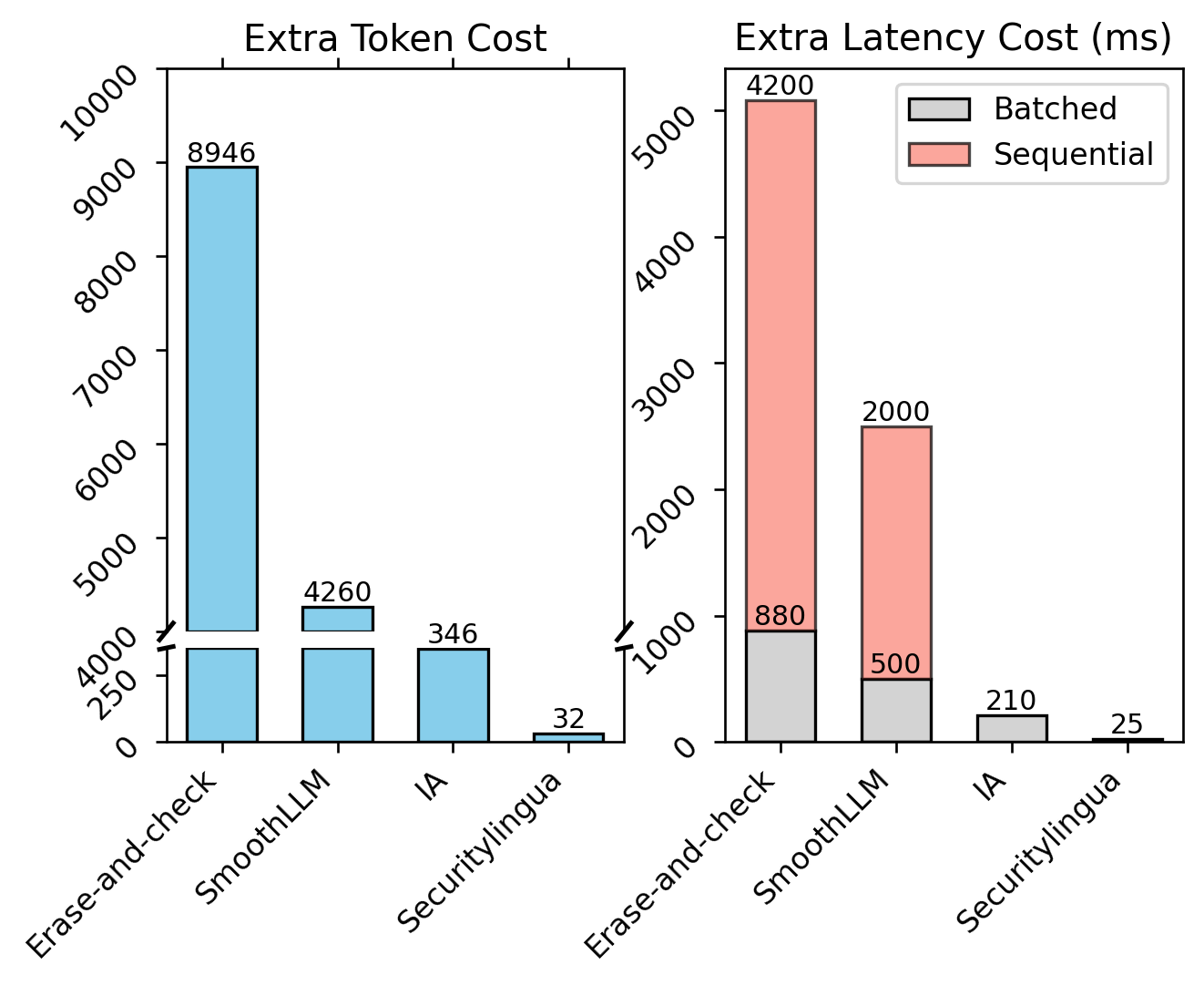}
    \caption{Cost comparison of SecurityLingua against our baselines. All  evaluations were conducted on a single A100 GPU with vLLM and the Transformers library.}
    \label{fig:cost}
\end{wrapfigure}

\textbf{Defense Efficiency.}
In Figure~\ref{fig:cost}, we show the extra latency and token cost incurred by various defense methods tested on JailbreakBench. As demonstrated, some of the defense methods are extremely expensive in terms of extra token cost. For example, SmoothLLM introduces 4,260 extra tokens to conduct extensive safety checks due to its permutation-then-check mechanism. Erase-and-check, on the other hand, incurs more than double the cost -- about 9,000 extra tokens on average are required to check each query. Based on the default setting reported in \citet{kumar2023certifying} and \citet{chao2024jailbreakbench}, Erase-and-check and SmoothLLM randomly sample 20 and 10 variations per query, respectively, which reduces their practicality for use in real production environments. In contrast, SecurityLingua incurs only 32 extra tokens on average, which is about 11\% of the original prompt length.

In terms of latency, SmoothLLM and Erase-and-check both incur significant costs, as they require models to generate answers for multiple variations of the prompt. Erase-and-check and SmoothLLM incur an extra 4,200 ms and 2,000 ms to finish the safety check process if executed sequentially, and 880 and 500 ms if executed in batched inference. In contrast, SecurityLingua incurs only 25 ms on average, as it only requires a single forward pass, and we can further reduce the latency in practice using batched inference. The IA method, although highly token-efficient compared to Erase-and-check and SmoothLLM, still incurs more tokens compared to SecurityLingua. IA makes additional LLM calls to check the query, which requires an auto-regressive decoding process and leads to much higher latency compared to SecurityLingua.

\begin{table}[t]
    \centering
    \resizebox{\textwidth}{!}{
    \begin{tabular}{lcccccccccc}
    \toprule
    & \multicolumn{2}{c}{ARC} & \multicolumn{2}{c}{GPQA} & \multicolumn{2}{c}{MMLU} & \multicolumn{2}{c}{GSM8K} & \multicolumn{2}{c}{Avg} \\
    \cmidrule(lr){2-3} \cmidrule(lr){4-5} \cmidrule(lr){6-7} \cmidrule(lr){8-9} \cmidrule(lr){10-11}
    Method & Acc. & Refusal (\%) & Acc. & Refusal (\%) & Acc. & Refusal (\%) & Acc. & Refusal (\%) & Score & Refusal (\%) \\
    \midrule
    None & 94.0 & - & 46.0 & - & 88.4 & - & 50.5 & - & 69.7 & - \\
    \midrule
    PPL Filter & 96.1 & 5.7 & 44.1 & 3.4 & 86.0 & 5.3 & 51.9 & 18.6 & 69.5 & 8.3 \\
    SmoothLLM & 84.1 & 4.7 & 39.2 & 2.9 & 70.2 & 8.6 & 38.7 & 0.2 & 58.0 & 4.1 \\
    Erase-and-check & 94.0 & 1.2 & 47.1 & 6.9 & 85.6 & 5.8 & 50.6 & 1.3 & 69.3 & 3.8 \\
    IA & 96.0 & 0 & 44.5 & 0 & 89.5 & 0 & 54.2 & 0 & 71.1 & 0 \\
    JDetector & 93.5 & 0 & 47.2 & 2.7 & 83.5 & 4.3 & 50.0 & 1.5 & 68.6& 2.1 \\
    SecurityLingua & 95.0 & 0 & 46.7 & 0 & 88.9 & 0 & 57.5 & 0 & 72.0 & 0 \\
    \bottomrule
    \end{tabular}
    }
    \caption{Comparison of GPT-4's performance on various tasks with and without defense methods. Acc. scores (higher is better) and Refusal rates (lower is better) are reported. SecurityLingua maintains a zero Refusal rate across all tasks and maintains, and slightly improves, accuracy.}
    \label{tab:utility_results}
\end{table} 

\textbf{Utility Test.}
In Table~\ref{tab:utility_results}, we show the impact of each defense method on model utility, i.e., the performance of the model on downstream tasks. First, we find that many defense methods will produce ``false positives''. For example, PPL Filter and SmoothLLM reject about 8\% and 4\% of queries, which can greatly diminish the overall utility of the system and user experience. SmoothLLM also leads to a notable performance drop on the benchmarks, from 69.7 to 58.0 in accuracy. This degradation is likely caused by the final answer being constructed from the outputs of multiple perturbed versions of the original query, where these perturbations may have introduced semantic inconsistencies.
In general, IA and SecurityLingua are more robust in terms of both performance and refusal rate, and on some benchmarks they even achieve better performance than the original model. This may be due to the fact that IA and SecurityLingua pre-process the query and enrich it with the request's true intention in advance of the eventual response generation.

\section{Conclusion}

We proposed SecurityLingua, a plug-and-play generic method to defend against jailbreak attacks. SecurityLingua employs prompt compression to efficiently detect malicious instructions and clarify user intent, enabling LLMs to identify attacks and generate safe outputs. We evaluate SecurityLingua along the dimensions of defense capability, efficiency and utility. The results show that SecurityLingua achieves strong defense performance with significantly lower latency and token cost compared to available baselines, while maintaining, and in some cases improving, the utility of the original model.

\clearpage

\bibliography{ref}
\bibliographystyle{colm2025_conference}

\appendix

\section{Appendix}
The prompts used in the our data annotation process.

\begin{figure}[htb]
    \begin{tcolorbox}[left=3pt,right=3pt,top=3pt,bottom=3pt]
    \small
    
    \textbf{Our Instruction for Compression}\\
    You are given a question and your task is to extract key words from the question to describe the central intention of the question. You should extract a continuous piece from the question to describe the central intention of the given question.
    \vspace{2mm}

    You should first analyze the question (up to 100 words) to understand the question and its intention. Then extract a continuous piece from the question to describe the central intention of the given question. Make sure your extracted piece is surrounded by \textcolor{blue}{\textless intention\textgreater} and \textcolor{blue}{\textless /intention\textgreater}. It's very important that your extracted piece appears literally in the given question.
    \vspace{2mm}

    Now extract key words from the following questions to reveal its intention: \textcolor{red}{\{QUESTIONS\}}.
    \end{tcolorbox}

    \caption{The prompt we use for the compression procedure with GPT-4o, Mistral-Large, and Uncensored-Llama2-72B.}
    \label{fig:compression_prompt}
\end{figure}

\begin{figure}[t]
    \begin{tcolorbox}[left=3pt,right=3pt,top=3pt,bottom=3pt]
    \small
    
    \textbf{Our Instruction for Extension}\\
    You are given a question and your task is to extend and rewrite the question with more context or in a more detailed manner. Feel free to use various contexts—professional, personal, imaginative, or informal—to make the revised question creative and diverse.
    \vspace{2mm}

    Please follow this format for each question:
    \vspace{2mm}

    You should: \\
    1) creatively expand the original question, adding context or details to make it more engaging and clear. \\
    2) your target length is \textcolor{red}{\{TARGET\_LENGTH\}} and you should produce a query \textcolor{red}{\{COMPLEXITY\}}.
    \vspace{2mm}

    Your output should be surrounded by \textcolor{blue}{\textless new\_question\textgreater} and \textcolor{blue}{\textless /new\_question\textgreater}.
    \vspace{2mm}

    Now do the task for the following questions: \textcolor{red}{\{QUESTIONS\}}.
    \end{tcolorbox}

    \caption{The prompt we use for the extension procedure.}
    \label{fig:extend_prompt}
\end{figure}

\end{document}